\PassOptionsToPackage{svgnames, x11names}{xcolor}
\documentclass[sigconf, natbib=true]{acmart}
\usepackage{csquotes} 
\usepackage{amsmath}
\usepackage{bm}
\usepackage{glossaries}
\usepackage[capitalise]{cleveref}
\usepackage{fontawesome5}
\usepackage{graphicx}
\usepackage{pgfplots}

\usetikzlibrary{external}
\crefname{equation}{Equation}{Equations}
\crefname{section}{Section}{Sections}
\crefname{figure}{Figure}{Figures}

\DeclareUnicodeCharacter{2212}{−}
\usepgfplotslibrary{groupplots,dateplot}
\usetikzlibrary{quotes, shapes,shapes.misc, graphs,patterns,shapes.arrows, arrows.meta, positioning, calc}

\newacronym[shortplural=LMs, longplural=Language Models]{lm}{LM}{Language Model}
\newacronym[shortplural=KGs, longplural=Knowledge Graphs]{kg}{KG}{Knowledge Graph}
\newacronym[shortplural=GUIs, longplural=Graphical User Interfaces]{gui}{GUI}{Graphical User Interface}
\newacronym{ged}{GED}{Graph Edit Distance}
\newacronym{onset}{OnSET}{Ontology and Semantic Exploration Toolkit}
\newacronym{nlp}{NLP}{Natural Language Processing}
\newacronym{ir}{IR}{Information Retrieval}
\newacronym{bto}{BTO}{Brainteaser Ontology}
\newacronym{mri}{MRI}{Magnetic Resonance Imaging}
\newacronym[shortplural=BGPs, longplural=Basic Graph Patterns]{bgp}{BGP}{Basic Graph Pattern}

\AtBeginDocument{%
  }

\setcopyright{acmlicensed}
\copyrightyear{2025}
\acmYear{2025}
\acmDOI{XXXXXXX.XXXXXXX}
\acmConference[SIGIR 2025]{The 48th International ACM SIGIR Conference on Research and Development in Information Retrieval}{July 13--18,
 2025}{Padua, IT}
\acmISBN{978-1-4503-XXXX-X/2018/06}

\begin{document}

\title{OnSET: Ontology and Semantic Exploration Toolkit}

\author{Benedikt Kantz}
\email{benedikt.kantz@tugraz.at}
\affiliation{%
  \institution{University of Technology Graz}
  \city{Graz}
  \country{Austria}
}
\author{Kevin Innerebner}
\email{innerebner@tugraz.at}
\affiliation{%
  \institution{University of Technology Graz}
  \city{Graz}
  \country{Austria}
}
\author{Peter Waldert}
\email{peter.waldert@tugraz.at}
\affiliation{%
  \institution{ University of Technology Graz}
  \city{Graz}
  \country{Austria}
}
\author{Stefan Lengauer}
\email{s.lengauer@tugraz.at}
\affiliation{%
  \institution{ University of Technology Graz}
  \city{Graz}
  \country{Austria}
}
\author{Elisabeth Lex}
\email{elisabeth.lex@tugraz.at}
\affiliation{%
  \institution{ University of Technology Graz}
  \city{Graz}
  \country{Austria}
}
\author{Tobias Schreck}
\email{tobias.schreck@tugraz.at}
\affiliation{%
  \institution{University of Technology Graz}
  \city{Graz}
  \country{Austria}
}

\renewcommand{\shortauthors}{Kantz et al.}

\begin{abstract}
  Retrieval over knowledge graphs is usually performed using dedicated, complex query languages like SPARQL. We propose a novel system, \gls{onset}, that allows novice users to quickly build queries with visual user guidance provided by topic modeling and semantic search throughout the application. \gls{onset} allows users without any prior information about the ontology or networked knowledge to start exploring topics of interest over knowledge graphs, including the retrieval and detailed exploration of prototypical sub-graphs and their instances. Existing systems either focus on direct graph explorations or do not foster further exploration of the result set. We, however, provide a node-based editor that can extend these missing properties of existing systems to support the search over big ontologies with sub-graph instances. Furthermore, \gls{onset} combines efficient and open platforms to deploy the system on commodity hardware.
\end{abstract}




\received{18 February 2025}

\maketitle
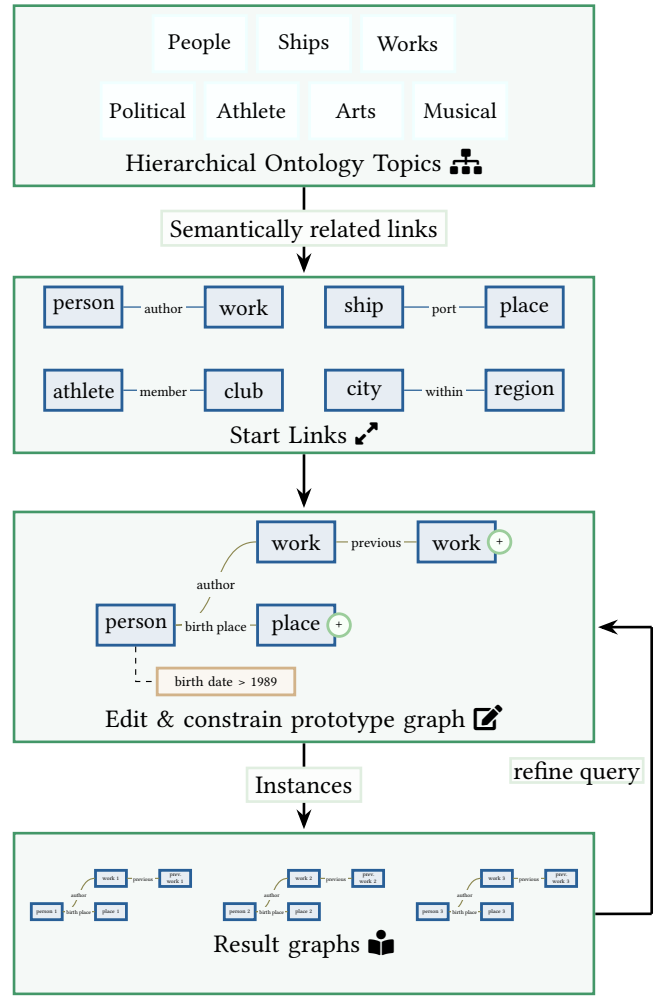
\begin{figure}[t!]
  \centering
  \vspace{6.2mm}
  \begin{tikzpicture}[
    scale=1.069, transform shape,
    mapping/.style={rectangle, draw=SeaGreen!90,  fill=SeaGreen!5,  text width=7cm, align=center, font={},  line width=1pt, minimum height=2cm},
    >={Stealth},
    every new ->/.style={shorten >=1pt, line width=1pt},
    every new --/.style={line width=1pt}
    ]
    \node (topics) [mapping] {
      \begin{tikzpicture}[
          topic/.style={rectangle, draw=Azure,  fill=Azure1!20, text width=0.95cm, align=center, font={\small},  line width=1pt, minimum height=0.7cm}
        ]
        \node (topic1)    [topic] {People};
        \coordinate (topic1_bel) [below=0.5cm of topic1];
        \node (topic1_1)  [topic, below right =0.1cm  and -0.55cm of topic1] {Athlete};
        \node (topic1_2)  [topic, below left =0.1cm  and -0.55cm of topic1] {Political};
        \node (topic2)    [topic, right =0.1cm of topic1] {Ships};
        \node (topic3)    [topic, right =0.1cm of topic2] {Works};
        \coordinate (topic3_bel) [below=0.5cm of topic3];
        \node (topic3_1)  [topic, below right =0.1cm  and -0.55cm of topic3] {Musical};
        \node (topic3_2)  [topic, below left =0.1cm and -0.55cm of topic3] {Arts};
      \end{tikzpicture}\\
      Hierarchical Ontology Topics \faSitemap
    };
    \node (start_graph) [mapping, below=1.1cm of topics]{
      \begin{tikzpicture}[
          linknode/.style={rectangle, draw=DodgerBlue4!90, fill=DodgerBlue4!10,  text width=0.75cm, align=center, font={\small},  line width=1pt, minimum height=0.5cm},
          tight/.style={draw=DodgerBlue4, font={\tiny}, inner sep=1pt},
          every edge quotes/.style={fill=SeaGreen!5, anchor=center, text depth=0cm, minimum height=0.3cm, anchor=mid, inner sep=1pt},
        ]
        \node (person) [linknode] {person};
        \node (work) [linknode, right=1cm of person] {work};
        \node (athlete) [linknode, below=0.5cm of person] {athlete};
        \node (club) [linknode, right=1cm of athlete] {club};
        \node (ship) [linknode, right=0.5cm of work] {ship};
        \node (place) [linknode, right=1cm of ship] {place};
        \node (city) [linknode, below=0.5cm of ship] {city};
        \node (region) [linknode, right=1cm of city] {region};
        \path (person) edge ["author", -, tight] (work);
        \path (athlete) edge ["member", -, tight] (club);
        \path (ship) edge ["port", -, tight] (place);
        \path (city) edge ["within", -, tight] (region);
      \end{tikzpicture}\\
      Start Links \faExpand*
    };
    \node (edit_graph) [mapping, below=0.7cm of start_graph]{
      \begin{tikzpicture}[
          linknode/.style={rectangle, draw=DodgerBlue4!90, fill=DodgerBlue4!10,  text width=0.75cm, align=center, font={\small},  line width=1pt, minimum height=0.5cm},
          constraint/.style={linknode, draw=Burlywood3!90, fill=Burlywood3!10,  align=center, font={\tiny},text width=0.6cm, minimum height=0.3cm},
          tight/.style={draw=Khaki4, font={\tiny}, inner sep=1pt},
          constraint_link/.style={dash pattern=on 2pt off 2pt},
          add_btn/.style={circle, fill=DarkSeaGreen1!10, draw=DarkSeaGreen3, line width=1pt, inner sep=1pt, minimum size=8pt,font=\tiny,anchor=center},
          every edge quotes/.style={fill=SeaGreen!5, anchor=center, text depth=0cm, minimum height=0.3cm, anchor=mid, inner sep=1pt},
        ]
        \node (person) [linknode] {person};
        \node (person_const) [constraint, below right=0.25cm and 0.25cm of person.south, text width=1.5cm] {birth date > 1989};
        \coordinate (bel_person) at (person_const.west -| person.south) [fill=red]{};
        \node (work) [linknode, above right=0.5cm and 1cm of person] {work};
        \node (work_next) [linknode, right=1cm of work] {work};
        \node (place) [linknode, right=1cm of person] {place};
        \node (add_place) at  ($(place.east)+(1pt,0)$) [add_btn] {+};
        \node (add_work) at  ($(work_next.east)+(1pt,0)$) [add_btn] {+};
        \path (person.east) edge ["author", -, tight, out=0,in=180] (work.west);
        \path (person.east) edge ["birth place", -, tight, out=0,in=180] (place.west);
        \path (work.east) edge ["previous", -, tight, out=0,in=180] (work_next.west);
        \draw[constraint_link] (person.south) -- (bel_person) -- (person_const.west);
      \end{tikzpicture}\\
      Edit \& constrain prototype graph \faEdit
    };

    \node (result_graph) [mapping, below=1.1cm of edit_graph]{
      \begin{tikzpicture}[
          scale=0.4, transform shape,
          linknode/.style={rectangle, draw=DodgerBlue4!90, fill=DodgerBlue4!10,  text width=0.75cm, align=center, font={\tiny},  line width=1pt, minimum height=0.5cm},
          constraint/.style={linknode, draw=Burlywood3!90, fill=Burlywood3!10,  align=center, font={\tiny},text width=0.6cm, minimum height=0.3cm},
          tight/.style={draw=Khaki4, font={\tiny}, inner sep=1pt},
          constraint_link/.style={dash pattern=on 2pt off 2pt},
          every edge quotes/.style={fill=SeaGreen!5, anchor=center, text depth=0cm, minimum height=0.3cm, anchor=mid, inner sep=1pt},
        ]
        \foreach \x/\i in {0/1,6/2,12/3}{
            \node (person) at ( \x cm, 0cm) [linknode] {person \i};
            \node (work) [linknode, above right=0.5cm and 1cm of person] {work \i};
            \node (work_next) [linknode, right=1cm of work] {prev. work \i};
            \node (place) [linknode, right=1cm of person] {place \i};
            \path (person.east) edge ["author", -, tight, out=0,in=180] (work.west);
            \path (person.east) edge ["birth place", -, tight, out=0,in=180] (place.west);
            \path (work.east) edge ["previous", -, tight, out=0,in=180] (work_next.west);
          }
      \end{tikzpicture}\\
      Result graphs \faBookReader
    };
    \node (right_edit) [right=0.7cm of edit_graph, radius=0pt] {};
    \node (right_res) [right=0.7cm of result_graph, radius=0pt] {};
    \graph[use existing nodes, edge quotes={draw=Honeydew2, anchor=mid, text depth=0cm, minimum height=0.3cm, line width=1pt, fill=Honeydew2!10 },]{
    topics->["Semantically related links"]start_graph->edit_graph->["Instances"]result_graph;
    };
    \graph[use existing nodes, edge quotes={draw=Honeydew2, anchor={east}, text depth=0cm, minimum height=0.3cm, line width=1pt, fill=Honeydew2!10 },]{
    result_graph--right_res.west--["refine query", inner sep=1pt, outer sep=3pt]right_edit.west->edit_graph;
    };
  \end{tikzpicture}
  \caption{\gls{onset} user flow. The user can select topics of interest and retrieve possible start links. These links are then expanded \& constrained within an editor, which finally retrieves different instances of the searched graph. A demo video is accessible at \url{https://cloud.tugraz.at/index.php/s/djdayXSSWAX4ajt}}
  \Description{The user flow, beginning with the broad range of topics contained within the ontology, like people, ships, athletes, and so forth. Each of these leads to start links, where entity classes are related to each other using links, like persons to their work. Selecting a start link opens the editor, where users can edit and append the prototype graph, which results in an instance graph.}
  \label{fig:small_teaser}
\end{figure}

\section{Introduction}
Information retrieval from knowledge graphs is usually performed using specialized languages like SPARQL~\cite{Seaborne2024Sparql, Gruber1995Onto}.

These query languages, however, require specialized knowledge of both the syntactical nature of the query language and the schema of the knowledge graph. This additional barrier of entry can be challenging to overcome for non-experts of these languages and first-time users, hindering easy exploration of these rich knowledge bases, like DBpedia~\cite{Lehmann2015DBpediaA}.

To further the use of ontologies and knowledge graphs by non-expert users, existing systems already provide natural language or visual interfaces, making these applications more accessible. We relate these works to ours in \cref{sec:related}. Compared to these existing systems, our system improves upon the initial exploration phase of the ontologies.

Our system, \gls{onset}, combines the existing interface paradigms to foster approachable exploration of ontologies. Our system considers notions from user guidance~\cite{Perez-Messina2023} to ease the exploration of large ontologies while keeping the system approachable.
None of the related works consider the missing knowledge of users that may need guidance towards a first information need. We solve this initial guidance problem using topic modeling over the ontology, an approach originating from \gls{nlp}. We furthermore extend the expansion of our prototype graph\footnote{The term \emph{example query} is used by related works \cite{vargas2019rdf, Lissandrini2020} to describe their approach of providing an instance from the knowledge graph. In contrast, our query paradigm builds on the \emph{prototype} for an instance.} with a semantic search interface to reduce the barrier of entry, as novice users do not have to have exact knowledge of the ontology, only semantic information derived from the topics presented initially. Furthermore, we expand the notion of dynamic results by providing instances of our prototype graph as small multiples, which can be inspected with further detail on demand. We utilize advances in \gls{nlp} as the basis for the search and exploration capabilities within \gls{onset}. Our application builds on open systems, starting with our SPARQL databases system qlever~\cite{Bast2017}, over the topic modeling approach based on BERTopic~\cite{grootendorst2022bertopic} and finally using open and efficient \glspl{lm}~\cite{teknium2024hermes3technicalreport,kusupati2024matryoshkarepresentationlearning}.


\section{Related Work}
\label{sec:related}

Existing retrieval systems already aim to reduce these barriers to entry. Some of these include natural language querying approaches~\cite{lei2018ontology, Lissandrini2020, Ferre2017SPARKLIS}, others build towards visual interfaces for building queries over knowledge bases~\cite{francart2023sparnatural, vargas2019rdf, LIU2022108870, GARCIA2022101235, Clemmer2011}.

Natural language interfaces focus on the ease of use and adaption of users' information needs as text, or, to some extent, incorporate \gls{nlp} concepts~\cite{lei2018ontology, Lissandrini2020}.  The \emph{NLP Engine}~\cite{lei2018ontology} aims to convert any natural language query to a SPARQL query for further processing using multiple translational layers. The system focuses on query mappings and assumes some prior knowledge of the schema users within the data to formulate queries in natural language. \emph{Graph Query Suggestion}~\cite{Lissandrini2020} expands the initial query graph of a user by suggesting related links using traditional \gls{ir} approaches. Furthermore, they build on the notion of example graphs, where the user builds a graph that serves as an example or prototype to retrieve similar structures from the knowledge graph.

On the other hand, the visual querying approaches focus on providing approachable systems to facilitate the use of knowledge graphs for non-experts. \emph{RDF Explorer}~\cite{vargas2019rdf} approaches this task by providing a graph editor, enabling the creation of graph examples to retrieve matching instances from the knowledge graph. The system provides users with query expansion options throughout the application, showing dynamic results during query building. The authors relate their work to previous query builders, notably \emph{Smeagol}~\cite{Clemmer2011}. This alternative follows similar exploration and retrieval paradigms but does not offer comparatively many SPARQL features. \emph{KGVQL}~\cite{LIU2022108870} defines a novel visual query language to ease the transformation between the visual querying and result set, at the cost of disregarding explorative approaches, favoring the proposed transformation approach to convert between data examples and queries. \emph{Rhizomer}~\cite{GARCIA2022101235} approaches the exploration of knowledge bases by providing different user interfaces to explore only at the top-level, graph-level, or only at the instance level of a single type. \emph{Sparnatural}~\cite{francart2023sparnatural} simplifies many of these explorative approaches into a tree-based approach that directly yields tabular results, limiting the easier result set exploration. Similarly, \emph{SPARKLIS}~\cite{Ferre2017SPARKLIS} approaches the topic by representing the query as a single natural language query expressed as a template over the ontology and combines the creation of the query with a small user interface system. Neither of these systems offers fuzzy search interfaces or initial user guidance to support novice users in their explorative search.


\section{Methodology}
\gls{onset} builds on task-driven user guidance design principles~\cite{Perez-Messina2023}. To this end, we first define the \emph{target} of the retrieval task to be a sub-graph of the ontology. This graph should be retrieved exploratively, i.e., the user can search for possible instances given some query. We aim to guide the user by \emph{leading} them towards subsets of interest and responding to user cues as they explore the ontologies.

We incorporate user guidance into the initial exploration step using BERTopic~\cite{grootendorst2022bertopic} to group the ontology into hierarchical topics and present the user with an overview of possible ontology aspects. Topic modeling, which is usually performed on text documents, is done by representing each class within the ontology as a single document using templates for the class, all its parent classes, and the properties associated with the class. The resulting topics are then labeled using a \gls{lm} to provide concise textual labels for each topic within the ontology. The novice user can, therefore, choose one or more topics of interest, which are then used to query for start links and classes. This retrieval of start elements uses the averaged semantic embedding of the selected topics and semantic embeddings of textual representations of all links and classes within the ontology.

\begin{figure}[tb]
  \centering

  \pgfdeclarelayer{background}
  \pgfsetlayers{background,main}
  \begin{tikzpicture}[
      anno_node/.style={
          circle, draw=Honeydew2, fill=Honeydew2!20, inner sep=2pt, radius=13pt
        },
    ]
    \begin{pgfonlayer}{background}
      \node (img) at (0cm,0cm) [anchor=south west, inner sep=0pt, ] {
        \includegraphics[width=\linewidth]{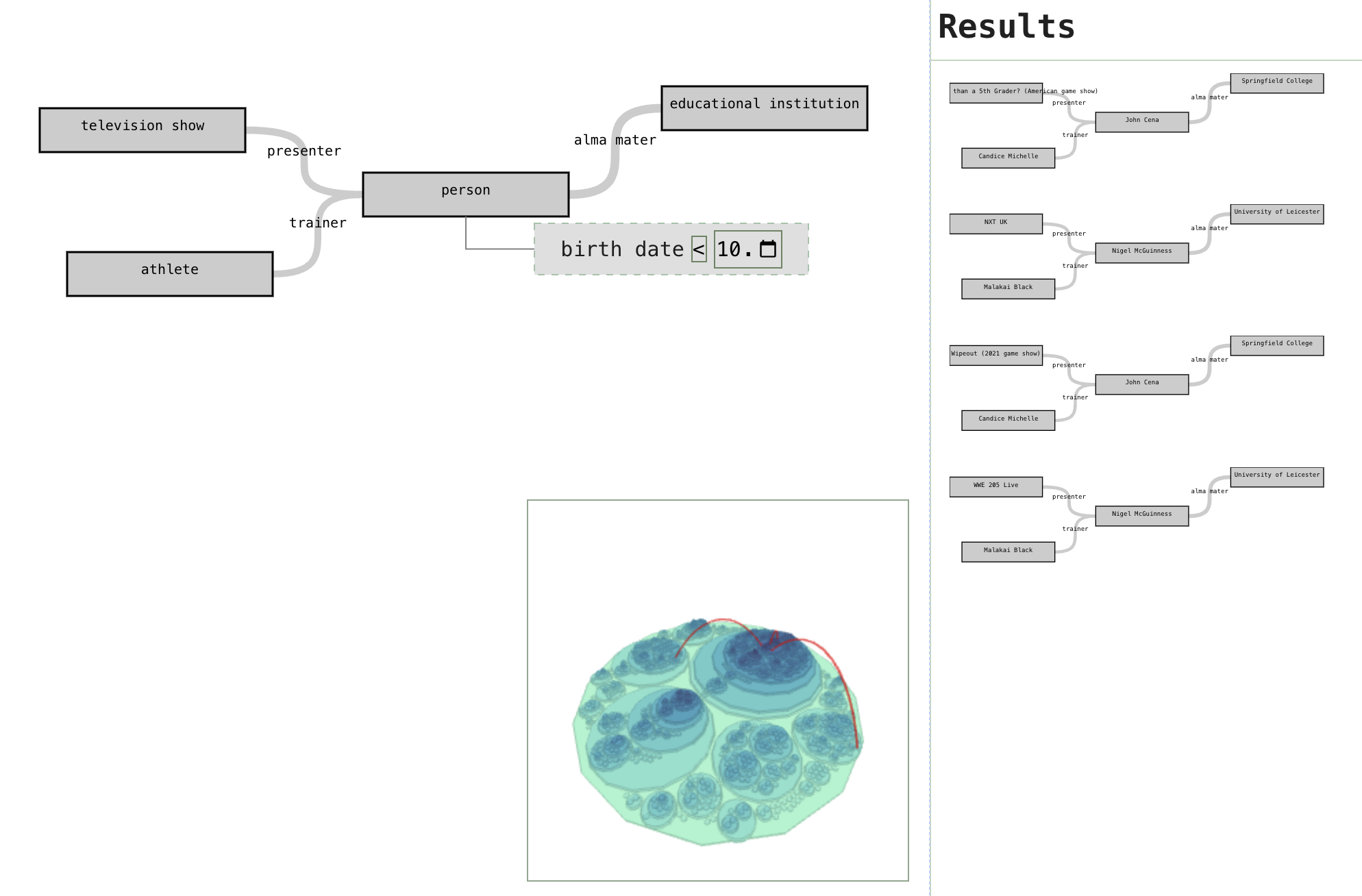}
      };
    \end{pgfonlayer}
    \draw (1cm,5.5cm) node (anno_3a)  [anno_node]  {(3a)};
    \draw (7.5cm,5.8cm) node (anno_3b)  [anno_node]  {(3b)};
    \draw (3.5cm,2.5cm) node (anno_3c)  [anno_node]  {(3c)};
  \end{tikzpicture}
  \caption{The query can be built using a straightforward interactive process. The user can add any allowed link within the ontology, with a visual indication of the prevalence within the knowledge graph by link width. Users can also add constraints to the nodes. (3a) The tool immediately provides visual results (3b) and provides a visual indicator of the explored classes and links using a small three-dimensional circle packing \enquote{minimap} in the bottom right (3c).}
  \Description{The interactive query editor, with the node editor in the top left, the result with similar instances to the right, and the overview minimap in the bottom.}
  \label{fig:query_build}
\end{figure}
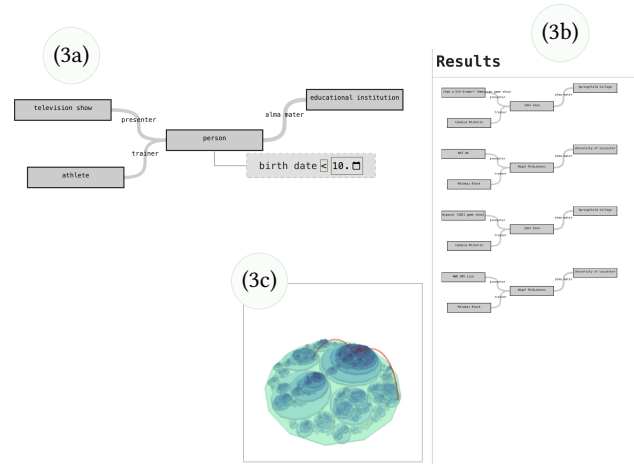

Once the user has chosen a start link based on the suggested set from the topic selection, the graph-building process is started. Our prototype graph-building system, shown in~\cref{fig:query_build}, is inspired by notions from prior work~\cite{vargas2019rdf,francart2023sparnatural} but extends them in crucial places. Our significant contribution is the inclusion of semantic search for all outgoing links between classes, enabling fuzzy search over the ontology. This guidance approach is more flexible than existing link expansion approaches~\cite{Lissandrini2020}, as it can respond to users' information needs in more flexible ways and is not bound to prior selections, giving the user more freedom while still adhering to the ontology. This semantic search is also applied to the attribute constraints, which can be searched similarly and applied to each entity. The notion of the prototype graph is closely related to the concept of \glspl{bgp}, with the addition of more constraints to the schema provided by the ontology and constraints placed upon properties by the user.

The resulting prototype graph (3a) is then used in three ways, similar to the levels of~\citet{GARCIA2022101235}, but integrated into the core retrieval process. First, the graph is used on top of a 3D circle-packing visualization of the ontology to show how the classes are distributed over the ontology with respect to its class hierarchy (3c). Second, the prototype graph is used to generate a SPARQL query to retrieve the intended instance set. This instance set populates the third use case of the graph, where we show small instances of the initial prototype graph. (3b). These instances, or result sets, can then be inspected, and the properties of the retrieved instances can be explored individually. The presentation of smaller, visually similar instances also differentiates our system from existing approaches, as we visually relate the result set and prototype graph.


\begin{figure}[tb]
  \centering
  \includegraphics[width=\linewidth, trim={1.3cm 13.5cm 1.8cm 4.5cm}, clip]{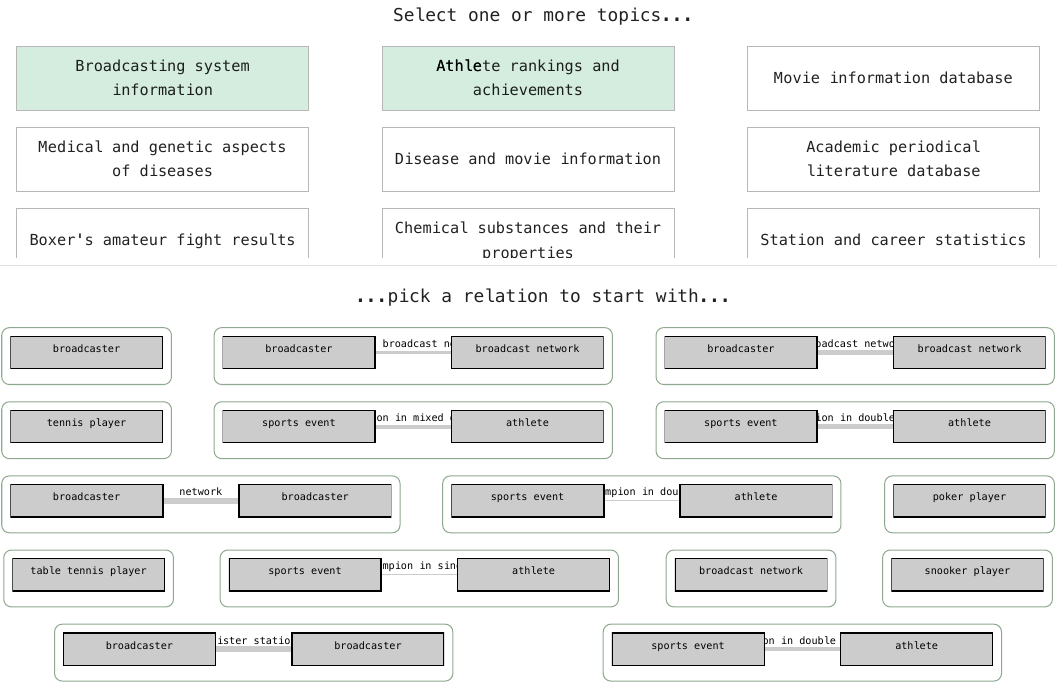};

  \caption{The initial selection process of \gls{onset} presents the user with a set of topics. The user can select one or more topics to start the exploration, select one starting link, and edit the graph in the next steps.}
  \Description{The interface in action shows some topics and the corresponding start links. The screenshot shows two selected topics, leading to a mixture of target links.}
  \label{fig:explore_topics}
\end{figure}

\section{Implementation}

The outlined concepts are integrated into our system, \gls{onset}, focusing on fast user responses even on larger ontologies. We furthermore base our whole stack on open-source systems to allow institutions or even users to start and tweak their systems.

To achieve our first goal of fast user responses, we only compute the topic modeling and embeddings of links and classes on the first startup with a specific ontology. We store our resulting hierarchical topic map and embeddings in PostgreSQL\footnote{\url{https://www.postgresql.org/}} with the help of the pgvector extension \footnote{\url{https://github.com/pgvector/pgvector}}, allowing fast retrieval given a query embedding. To generate these embeddings quickly, even on commodity hardware, we use the \verb|stella_en_400M_v5|\footnote{\url{https://huggingface.co/NovaSearch/stella_en_400M_v5}} model, which is, at the time of submission, the best-performing smaller model w.r.t. the massive text embedding benchmark~\cite{muennighoff2022mteb}. The precomputed topics are generated using the same embedding model, and additional topic labels are generated using the Hermes Llama 3.2 8B model~\cite{teknium2024hermes3technicalreport}.

While these efforts improve the query time for the prototype graph building, live updates of the result set require a similarly fast system. To achieve those fast responses, even on more complex queries and on commodity hardware, we use qlever~\cite{Bast2017}, a SPARQL query engine that outperforms most existing engines in both speed and system requirements. This speedup enables our system to serve and display updates as the user builds their query, aiding the user in retrieving non-empty sets and showing intermediate results to guide the search even further.

Our user interface builds on Vue.js\footnote{\url{https://vuejs.org/}}, in combination with three.js\footnote{\url{https://threejs.org/}} and D3.js~\cite{2011-d3}. All the used database systems and models are open-source and open-weight, providing state-of-the-art performance in their respective fields while still being able to run on commodity hardware.


\section{Case study}

We present \gls{onset} in the scope of two case studies for explorative search over two different ontologies. First, we show how a novice user might start their exploration of DBpedia~\cite{Lehmann2015DBpediaA} to discover interesting facts and relations. Our second use case covers the \gls{bto}~\cite{Faggioli2024} and how experts within a field might approach more specialized ontologies.

\subsection{Exploring DBpedia}

DBpedia~\cite{Lehmann2015DBpediaA} is a curated ontology and knowledge graph derived from Wikipedia, providing general linked information. Querying and exploring this knowledge base usually requires the SPARQL language. Our visual exploration toolkit, \gls{onset}, enables the novice user to start exploring the knowledge base immediately through the presented topics.

An example of such an exploration flow could be the initial selection of the topic \enquote{Broadcasting system information} and \enquote{Athlete rankings and achievements}, as seen in \cref{fig:explore_topics}. This selection queries our system for links similar to the selected topic and displays start link suggestions to the user. The user can, therefore, start exploring DBpedia without prior knowledge of the classes and links contained in it, hinting the user at possible links within the ontology. Next, the user chooses a link out of the suggested list, starting the building process of the prototype graph. The user then adds links and constraints to the prototype graph, drilling down towards a specific result set of interest---in this case, persons who trained an athlete and presented a television show and where they were educated. \gls{onset} guides the user towards non-empty sets by indicating prevalence in the knowledge graph through link strength. The user can assume that the result set is not empty due to the strong links between all classes on the prototype graph, which can be immediately verified as the result set is updated directly. While building the graph, the \enquote{minimap} in the corner of \cref{fig:query_build} is dynamically updated to indicate which regions within the class hierarchy are covered. In this case, two links span the class hierarchy while one link, the athlete's trainer, covers only the local hierarchy within the person region.


\subsection{Exploring \gls{bto}}

Our second example outlines the explorative search over the \gls{bto}~\cite{Faggioli2024}, an ontology and knowledge graph containing semantic knowledge about patients, caregivers, and their diagnoses. The explorative process for an interested party starts similar to the use case above but is presented with different topics.

\begin{figure}[tb]
  \centering
  \includegraphics[width=0.5\linewidth]{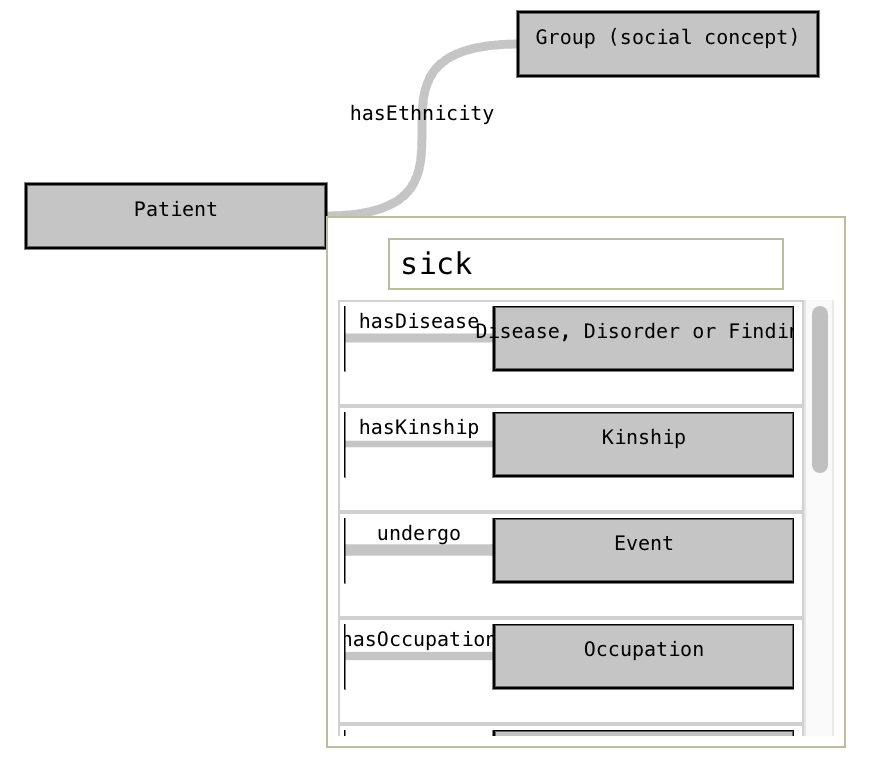}
  \caption{The search interface to retrieve semantically similar links, on the example of \gls{bto}~\cite{Faggioli2024}.}
  \Description{The popup is shown to the user when they want to add new links. The user can perform a semantic search over all allowed links, with an indicator of the prevalence within the \gls{kg} shown through the width of the link.}
  \label{fig:bto_semantic_search}
\end{figure}

The user might be interested in diseases associated with certain relations, so they start their search with the topic \enquote{Kinship types and relations}, starting with the link \enquote{hasEthnicity}. This single relation is too general, so they search for sicknesses, as seen in \cref{fig:bto_semantic_search}. Using the semantic search capabilities, they find the relation \enquote{hasDisease} within the ontology and add them to their prototype graph. Nevertheless, this relatively small prototype graph allows the user to compare and search over these links of interest in the result set. 

\section{Future Work}

\gls{onset} provides a low barrier of entry for novice users to knowledge graph exploration. We intend to improve upon the breadth of queries the user can express through our system in the future. A current limitation of our system is the inability to specify complete graphs, as the user can, for the time being, only build tree-like graphs, while closed graphs might be of interest for more advanced or intricate use cases.

We intend to refine the constraint application process as the filtering strength of the properties of an instance is not clearly visually defined yet, which could aid the user in exploring and retrieving data more attuned to their need and assist in non-empty result retrieval. Another interesting avenue is the extension towards optional parts of the prototype graph, both in the form of links and constraints, to allow more fuzzy retrieval. Another missing aspect of the constraint-building process is missing properties, where a search over multimodal data types like spatial or image data could be interesting.

Other visual cues to ease the query-building process could be of interest like result set changes upon adding or removing query parameters as a what-if visual cue in all affected results and \enquote{minimap}.

\section{Conclusion}

The presented \gls{ir} system, \gls{onset}, allows novice users without any prior information about the knowledge base to build queries and explore it in an integrated way. We first present related systems that have a similar aim of allowing non-expert users to build SPARQL queries and examine knowledge graphs. Our approach, however, differs through the use of initial guidance approaches to lower the barrier of entry even further by providing semantic search for both the initial link search and the prototype graph expansion, and, finally, immediate result feedback built right into the interface. We, furthermore, display an overview of the ontology to relate the current query to the whole ontology.

We emphasize using specialized open database systems to provide fast response time, enabling interactive and immediate result exploration even for minor changes to the prototype graph. We finally demonstrate two possible explorative user flows using \gls{onset} to inspire further use cases.


\begin{acks}
  This work is partially supported by the HEREDITARY Project, as part of the European Union's Horizon Europe research and innovation programme under grant agreement No GA 101137074, the Austrian Science Fund (FWF) 10.55776/COE12, Cluster of Excellence {\href{https://www.bilateral-ai.net/home}{Bilateral Artificial Intelligence}} and the FFG HybridAir project \#FO999902654. 
\end{acks}

\bibliographystyle{ACM-Reference-Format}
\bibliography{references}

\appendix









\end{document}